\documentclass[aps,prl,twocolumn,superscriptaddress,longbibliography,groupedaddress]{revtex4}
\usepackage{graphicx}  
\usepackage{dcolumn,comment}   
\usepackage{bm}        
\usepackage{amsmath}   

\usepackage{color}

\begin{document}

\title{Self-propelled evolution on regenerating landscapes}
\author{Alexander Heyde}
\affiliation{Department of Organismic and Evolutionary Biology, Harvard University, Cambridge, MA 02138}
\affiliation{Department of Applied Physics, Stanford University, Stanford, CA 94305}
\author{L.~Mahadevan}
\affiliation{Department of Organismic and Evolutionary Biology, Harvard University, Cambridge, MA 02138}
\affiliation{School of Engineering and Applied Sciences, Harvard University, Cambridge, MA 02138}
\affiliation{Department of Physics, Harvard University, Cambridge, MA 02138}
\date{\today}

\begin{abstract}
\noindent Evolving populations both respond to and reshape their environments, making fitness landscapes dynamic rather than static. We present a minimal eco-evolutionary model that couples replicator dynamics for a population density with a regenerating resource-driven landscape through a single environmental sensitivity parameter. This allows evolving populations to generate and ride self-induced selection gradients, enabling directed motion in trait space even on initially flat landscapes. Our analysis reveals sustained oscillations, chaotic dynamics, and evolutionary branching. To explain these, we derive reduced dynamical equations that extend Fisher’s fundamental theorem to deformable landscapes by incorporating curvature-driven variance dynamics and environmental feedback. Together, these results show how populations actively reshape and self-propel themselves on regenerating landscapes.


\end{abstract}

\maketitle

Evolution changes the distribution of traits in a population through differences in reproductive success. Standard treatments describe this process as movement on a fixed fitness landscape, where selection drives populations toward locally optimal states \citep{wright1932roles, reeve1993adaptation,orzack2001adaptationism}. However, organisms also modify their environments. Populations consume resources, alter ecological structure, and influence the very conditions that determine their fitness \citep{lewontin2000evolutionary,odling1996niche,day2003rethinking}. These changes can occur on a vast range of timescales, creating feedback between evolutionary change and environmental transformation \citep{lewontin2000evolutionary,odling1996niche,day2003rethinking}. Such feedback is implicit in niche construction, coevolution, and eco-evolutionary dynamics, yet most theoretical frameworks treat the environment as externally specified, slowly varying, or independent of population composition. Here we present a minimal model to describe this fundamental feedback loop between adaptation and environmental change, drawing on formalisms from replicator dynamics \citep{smith1982evolution,hofbauer1998evolutionary,nowak2006evolutionary}, adaptive dynamics \citep{metz1995adaptive,mcgill2007evolutionary,fisher2021}, and ecological and evolutionary dynamics \citep{gurney1998ecological,volterra1928variations,Mahadevan2023,krishnan2026}.

We define the population density $n({\bf x},t)$ as a distribution over points $\mathbf x$ in trait space $X$ that varies over time $t$, and  $u({\bf x},t)$ to be an environmental variable also associated with each possible trait and time. A point $\mathbf x$ represents an ordered set of heritable features, and for tractability, we consider large populations so as to eliminate the role of demographic stochasticity and consider perfectly heritable traits. We assume that both density fields remain normalized at all times, i.e. $\int_X n({\bf x},t)d{\bf x} = \int_X u({\bf x},t) d{\bf x} = 1 $, but can be redistributed dynamically through trait space to drive the evolutionary process \footnote{Allowing for a non-conserved environment would lead to nonlinear temporal rescaling of our dynamics which would alter the rate of evolution.}.  We allow the population density $n({\bf x},t)$ and environmental landscape $u({\bf x},t)$ to follow the dynamical equations
\begin{align}
	\dot n &= n(u-\phi), \\
	\dot u &= -\varepsilon u(n-\phi),
\end{align}
where $\phi (t)=\langle n,u\rangle=\int_X n({\bf x},t) u({\bf x},t) \textup d\mathbf x$ denotes the average fitness of the population. Equation (1) is the usual replicator equation of evolutionary dynamics \citep{nowak2006evolutionary} and implies that individuals with higher than average fitness ($u>\phi$) will outcompete those with lower than average fitness ($u<\phi$).  Equation (2) for the landscape dynamics is written analogously with a rate of change (relative to the population) controlled by $\varepsilon$. Our model implicitly assumes that interactions are either perfectly local (e.g.\ the terms $nu$) or perfectly global (e.g.\ the terms of the form $\phi n, \phi u$): individuals occupying different points in trait space compete indirectly via the average fitness $\phi$ irrespective of their distance \footnote{This is an important limitation of our model; in general, these interactions may be complex and lead to spatial gradients, a problem that we will consider in the future.}.
For $\varepsilon>0$,  the landscape with higher-than-average occupancy ($n>\phi$) will tend to become less fit over time due to crowding and competition, while locations with lower-than-average occupancy ($n<\phi$) will become fitter in comparison. For $\varepsilon>1$, the environment is more sensitive to the population than the reverse, while for $\varepsilon<1$, the opposite is true. We refer to $\varepsilon$ as the relative sensitivity of the environment or landscape, and the coupled system represents the gradual shaping of the environment by a population evolving within it, and in turn the influence of the changing environment on the population  (Fig.\ 1).

When $\varepsilon=0$, fitness $u=u(\mathbf x)$ is a function only of trait space, and it immediately follows that the average fitness $\phi$ must be non-decreasing, such that selection is always adaptive, well known as the Lande equation \citep{fisher1958genetical,frank1992fisher,ewens1989interpretation,li1967fundamental}; indeed,  $\dot\phi=\int_X \dot nu\, \textup d\mathbf x =\langle n,(u-\phi)^2\rangle \equiv \sigma^2 $ is non-negative and represents the variance in fitness \footnote{Since $u^2-\phi u = (u-\phi)^2+\phi (u-\phi)$, $\int \dot n u \mathbf dx = \int n(u-\phi)^2 \mathbf dx+ \int n\phi(u-\phi) \mathbf dx= \phi \int nu \mathbf dx - \phi \int n \mathbf dx = \phi - \phi = 0$.}. However, when $\varepsilon\neq0$, we find instead that $\dot\phi=\sigma^2-\varepsilon\omega^2$, where $\omega^2=\langle(n-\phi)^2,u\rangle$ represents the variance in occupation among resources on the landscape (see SI). Our expression for $\dot\phi$ is related to the Price equation \citep{price1970selection,frank2012natural}, which in its continuous time form describes the dynamics of the mean $p$ of a trait of interest, $\frac{\textup d}{\textup dt} \mathtt{E}[p]=\mathtt{Cov}[u,p]+\mathtt{E}[\dot p],$
where the covariance and expectation are both taken with respect to the population density $n$. If the trait of interest is fitness itself, i.e.\ $p=u$ and hence $\mathtt{E}[p]=\phi$,  these correspond directly to the two terms in our expression for $\dot \phi$, a result well known in population genetics~\cite{nowak2006evolutionary}.

\begin{figure}
	\centering  
	\includegraphics[width=1.0\linewidth]{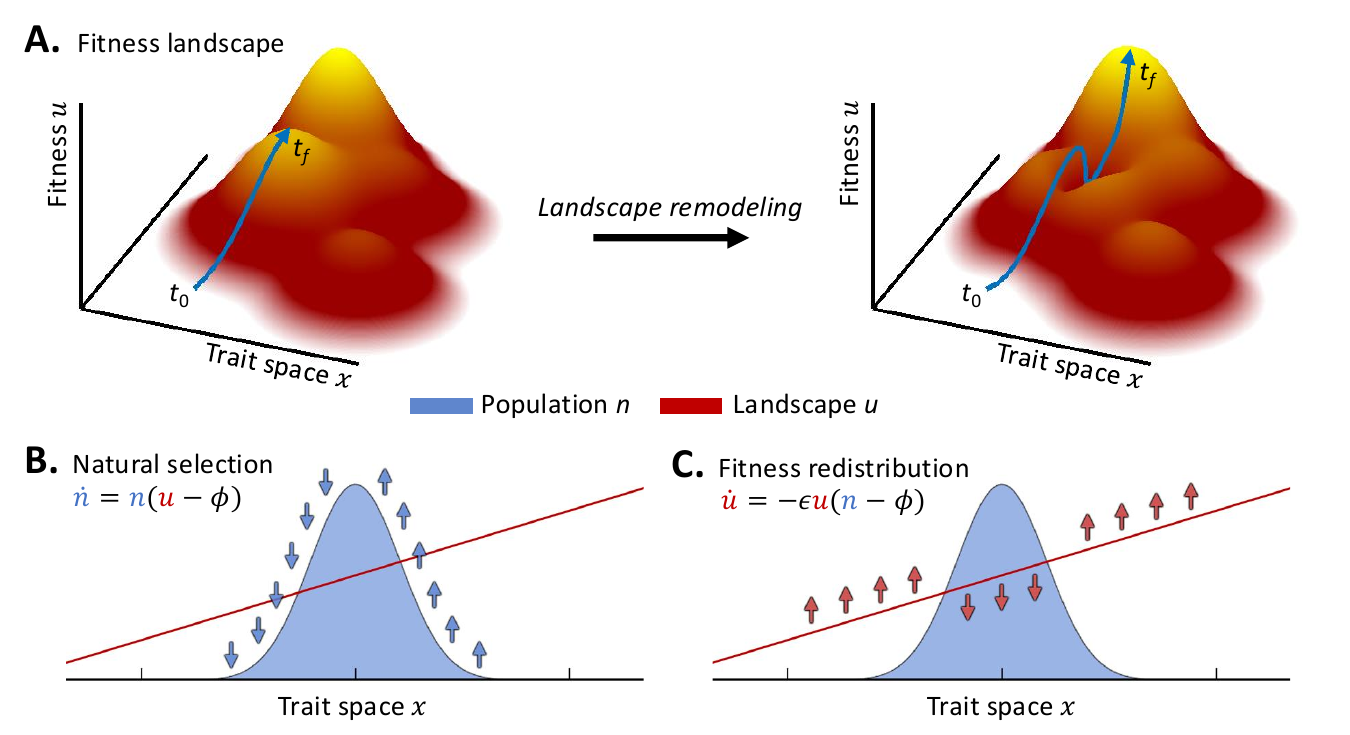}
	\caption{\textbf{Adaptation and landscape remodeling on a regenerating fitness landscape.} (A) In the traditional framework of a static fitness landscape (red, left), a population adapts and moves through trait space (blue arrow) by climbing local selection gradients toward a local fitness peak. In contrast, on a regenerating fitness landscape in which the population depletes local resources to reshape and generate new selection gradients, a population can propel away from a local peak to continue adapting over time (right). (B) On a one-dimensional landscape (red line) with constant slope, natural selection drives the population (blue) up the gradient via differential reproduction and death (blue arrows) as in Eq.\ 1, while resource depletion and regeneration alters the landscape over time (red arrows) when $\varepsilon>0$ as in Eq.\ 2.}
\end{figure}

\begin{figure*} 
	\centering
	\includegraphics[width=\linewidth]{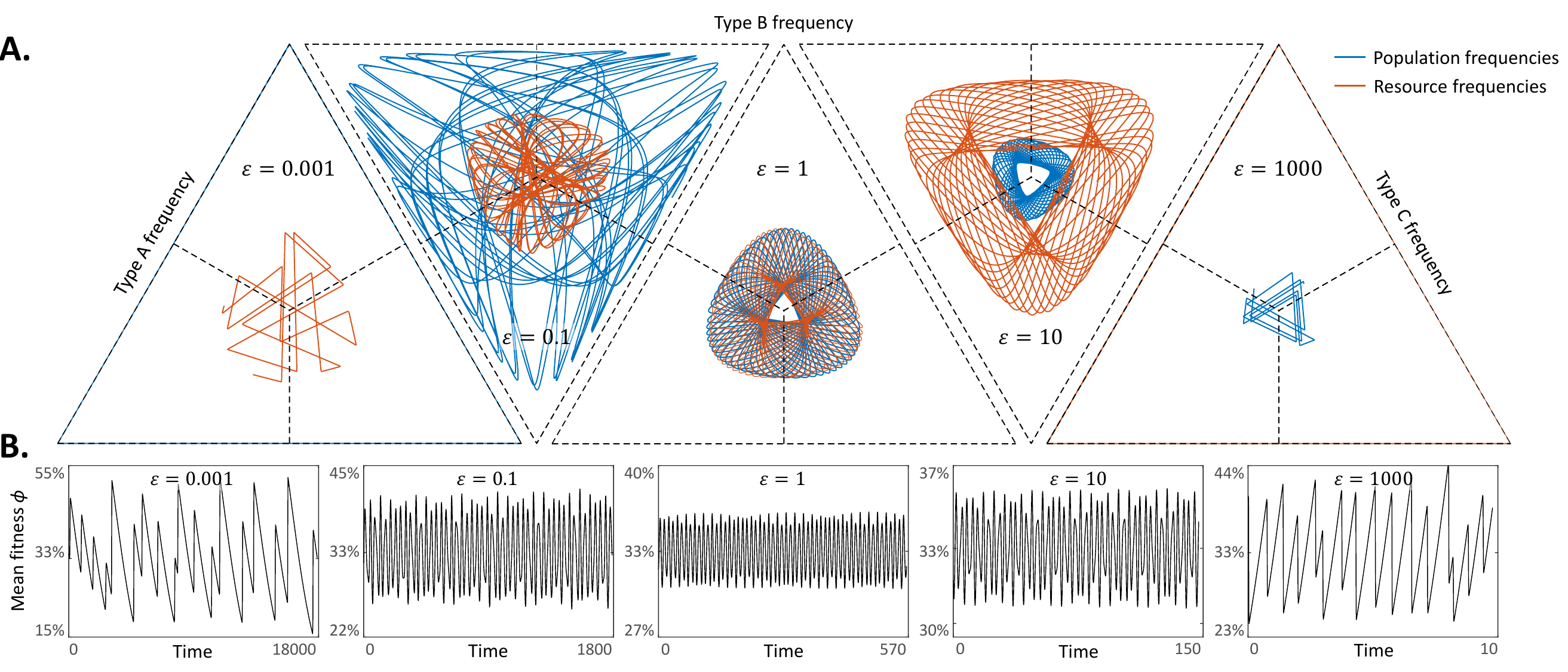}
	\caption{\textbf{Evolutionary dynamics of disjoint traits as the landscape sensitivity $\varepsilon$ varies}. Here we depict the dynamics for a simple trait space of only three points, representing three distinct phenotypes (types A, B, and C). (A) The population and landscape frequencies $n(\mathbf{x}_i)$ and $u(\mathbf{x}_i)$ are normalized and thus each can be mapped onto a point on a triangular simplex that follows Eqs.\ (1-2), revealing the dynamics of the population (blue) and landscape (red) distributions. For $\varepsilon=0.001$ (left), the population evolves quickly relative to the landscape and is almost always dominated by a single type (simplex vertices); in contrast, for $\varepsilon=1000$ (right), the most fit trait state switches rapidly, and the population drifts slowly towards this state in between transitions. Chaotic oscillations span the space in between these two extremes, with ordered symmetric dynamics near $\varepsilon=1$ (center). (B) For each simplex plot and $\varepsilon$ value in (A), we plot the mean fitness $\phi$ over time. Because $\varepsilon>0$, the mean fitness is not non-decreasing but instead fluctuates over time as the landscape changes. }
\end{figure*} 



We start by briefly considering the simplest case of only two types in a population, with the trait set $\mathbf x=\mathbf x_1, \mathbf x_2$.  In the traditional case of a static landscape ($\varepsilon=0$), the fittest type will come to dominate the entire population; yet if $\varepsilon>0$, fitness is continuously redistributed from the more prevalent type to the less prevalent type, resulting in oscillations in $n$ and $u$ that follow level-set trajectories of a conserved quantity $\kappa=\int_X \left[\varepsilon\log n+\log u\right]\,\textup{d}\mathbf{x}$, this known conserved quantity is associated with the magnitude of species fluctuations \citep{Pearce2020}.

Extending the analysis to three or more types introduces the potential for more complicated behavior, which we study as coevolving population and landscape trajectories on the simplex (Fig.\ 2). For small positive $\varepsilon\ll1$ (Fig.\ 2A left), the population evolves very quickly relative to the landscape and hence is almost always dominated by a single type (simplex vertex), only occasionally switching from one type to another (simplex edge) after sufficient change in the landscape has accumulated. In contrast, for large positive $\varepsilon\gg1$ (Fig.\ 2A right), the most fit trait state switches rapidly, and the population drifts slowly towards this state in between transitions. Closed and chaotic oscillations span the space in between these two extremes, with symmetrically ordered dynamics at $\varepsilon=1$ (Fig.\ 2A center). We emphasize that, for any $\varepsilon\neq0$, the mean fitness $\phi$ will not be non-decreasing but instead fluctuate over time as the landscape changes (Fig.\ 2B).
	


Next we consider a continuously variable trait with an initially perfect linear correlation to fitness, such that the trait is itself a measure of fitness, $u(x,0)=x$. If $\varepsilon=0$, the landscape is static, and $\dot n=n(x-\phi)$. As Fig.\ 3A depicts, if a population's fitness is initially normally distributed on this landscape with variance $\Sigma$, it will remain so over time with no change in the variance; however the mean fitness $\phi$ will increase linearly at a rate proportional to that variance in accordance with the Lande equation, resulting in a population that leverages its diversity to climb the selection gradient. Then the population will evolve in the direction of increasing fitness, i.e.\  $\dot n=-\nabla\cdot(\Sigma n\nabla u)$, a traveling pulse climbing towards ever-larger fitness values (see SI). Specifically, for $u=x$, the only profile that satisfies both the replicator equation (1) and the taxis equation is the Gaussian profile
\begin{align}
	n(x,t)=\frac{1}{\sqrt{2\pi\Sigma}}\exp\left\{-\frac{(x-\Sigma t)^2}{2\Sigma}\right\}.
\end{align}

In contrast, in the regime where $\varepsilon \ll \Sigma^{1/2}$, i.e.\ large relative to the population width, the landscape is effectively flat in the neighborhood of the population. In this limit of neutral evolution, $u=\phi$ everywhere and hence $\dot n=0$, resulting in no evolutionary selection over time, at least in the absence of stochastic features like genetic drift. And yet, as long as $\varepsilon>0$ so that $u$ changes in time, then $u=\phi$ is only necessarily true at $t=0$. 
To illustrate this, we consider a population initially distributed normally on a flat one-dimensional fitness landscape (Fig.\ 3B). Over time, the population will consume resources, locally depleting $u$ and thereby inducing a selection gradient in either  direction. The population will then proceed to move up this newly induced selection gradient, and as it travels, it continues to deplete resources. This process continues as the population is pulled by the fitness landscape up the selection gradient, and simultaneously pushes the fitness landscape in the same direction, ensuring there remains a local gradient to climb at all times.

\begin{figure}[!t]
	\centering
	\hspace*{-.13in} \includegraphics[width=1.07\linewidth]{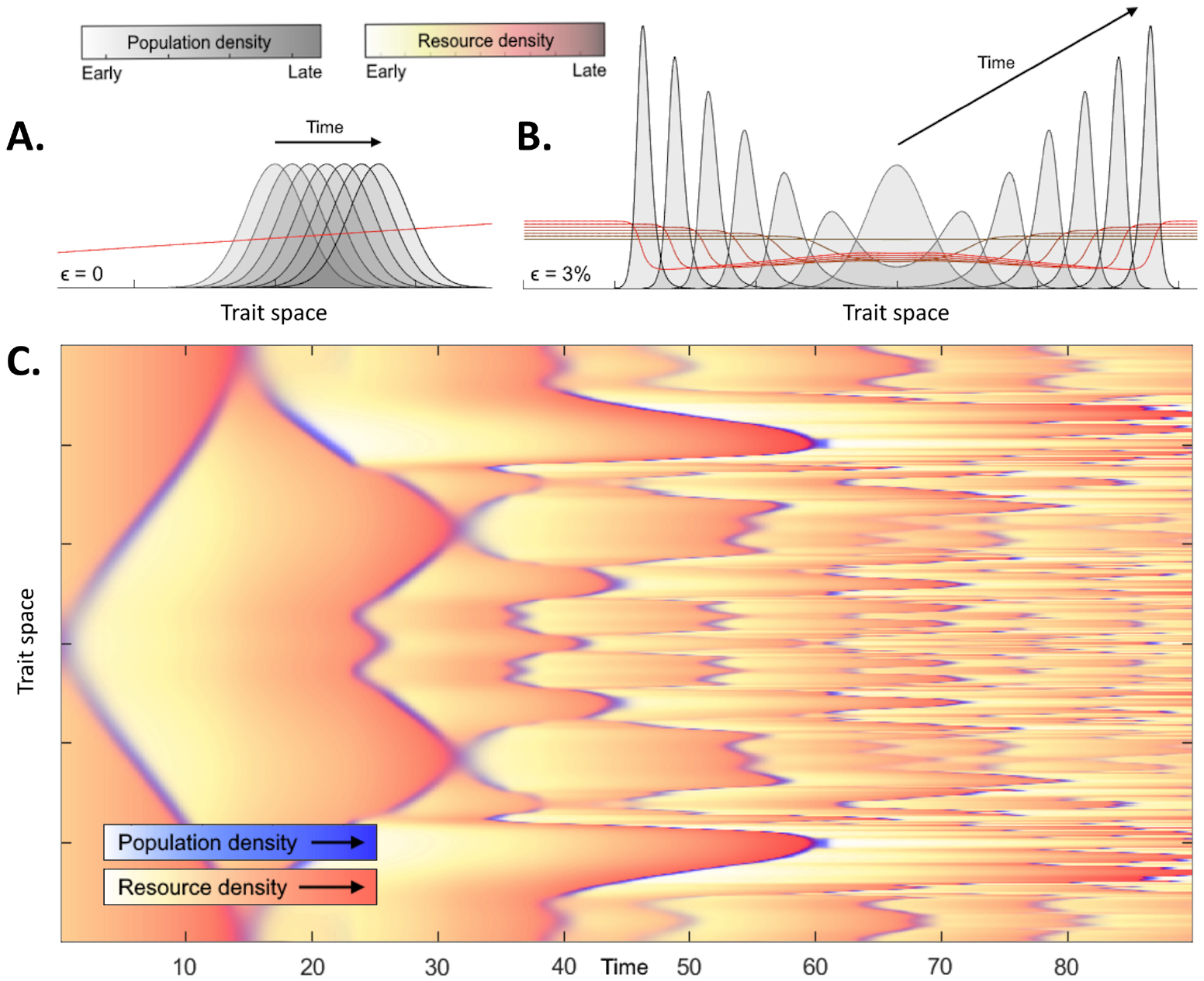}
	\caption{\textbf{Landscape sensitivity $\varepsilon$ affects evolution and speciation.} 
    (A–B) An initially Gaussian population evolves on a one-dimensional landscape $u \sim 1 + \beta x$ under Eqs. (1–2). (A) When the landscape is static ($\varepsilon = 0$), the population moves up the selection gradient at a rate that scales with $\beta>0$. (B) When the landscape is dynamic ($\varepsilon = 0.03$), even on an initially flat landscape ($\beta=0$), the population generates new selection gradients and self-propels, splitting into two traveling waves. (C) A kymograph shows repeated branching and evolution on a dynamic landscape for $\varepsilon = 0.03$; blue and red opacity denote population and resource density. The number of population modes fluctuates but grows on average, with a speciation rate scaling as $\varepsilon^{1/2}$ (see End Matter).}
\end{figure}

The magnitude of self-propelled evolution on a regenerating landscape varies with $\varepsilon$: the more sensitive the environment is to the population, the stronger the selection gradient and the stronger the positive feedback loop becomes, until the effect saturates at high $\varepsilon$. One metric to test this hypothesis is the rate at which the population moves through trait space on an initially flat landscape, i.e.\ the rate at which the mean trait value $\hat x$ changes in time, $R=\dot{\hat x}$. In Figure S2A, we show that for $\varepsilon$ values below some threshold $\varepsilon^*\approx0.4$, the rate of evolution increases with $\varepsilon$, and for a wide range of values converges to a square root scaling $\dot{\hat x}\sim\varepsilon^{1/2}$. However, as $\varepsilon$ increases beyond $\varepsilon^*$, the environment would be so sensitive that depletion would cause the local selection gradient to become dramatically more shallow, as $u\to0$ in the presence of the population; this rapidly slows evolution to a regime in which greater values of $\varepsilon$ are associated with decreased evolutionary rates. 
For this reason, a population that seeks to evolve most rapidly should seek to tune the magnitude of its impact on the environment such that $\varepsilon$ is neither too low nor too high. This pattern holds also on a landscape that is initially flat but noisy, sloped, or both sloped and noisy (Fig.\ S2A), with the only changes happening for very small $\varepsilon$ (for which the initial landscape profile persists for long times). 

\begin{figure} 
	\centering
	\includegraphics[width=\linewidth]{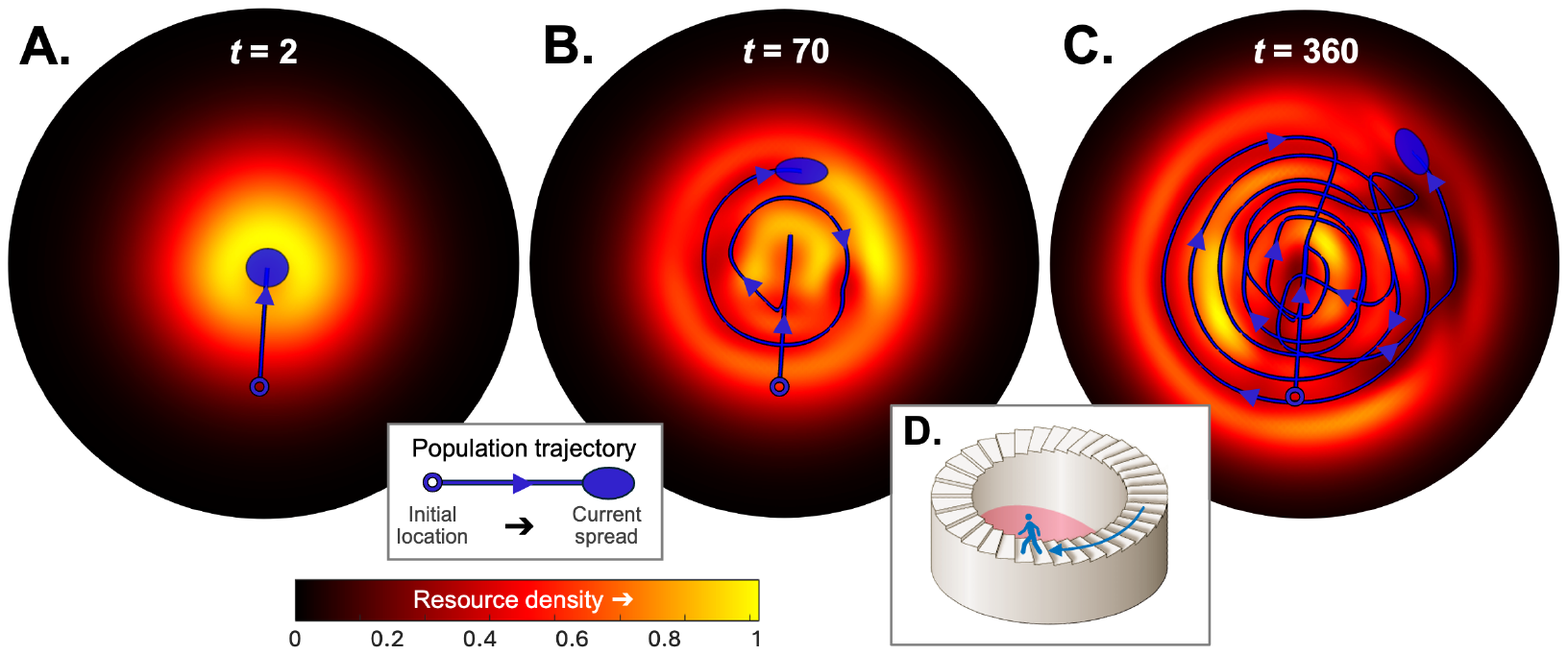}
	\caption{\textbf{Adaptive dynamics in a two-dimensional trait space show self-propelled oscillations.} A Gaussian population evolves on a broader initially Gaussian landscape (color scale) with $\varepsilon = 0.001$ under Eqs.\ (1–2). The blue line shows the trajectory of the population mean from its initial location (open blue circle) to a later location (blue ellipse:\ ±1 SD). (A) Early ascent toward the landscape peak from $t=0$ to $t=2$. (B) By $t=70$, the landscape has been reshaped to create new selection gradients that drive outward cyclic motion. (C) By $t=360$, long-term motion shows repeated loops and chaotic direction reversals as the landscape is repeatedly remodeled. (D) Schematic analogy:\ evolutionary adapation via continual climbing of local gradients on a landscape altered by the population, similar to climbing a closed-loop Penrose staircase.}
    \end{figure}

The feedback loop in our model not only accelerates the rate of evolution, but it is sufficient to produce evolutionary branching, whereby a population mode corresponding to a local density maximum can bifurcate and increase the population modality over time (Fig.\ S2B). If each mode is interpreted as a distinct ecotype or species, this can be thought of as a speciation rate; the number of modes grows exponentially with a speciation rate that scales as $\varepsilon^{1/2}$ (Fig.\ S2B inset and End Matter). An example of this speciation effect on a wrapped one-dimensional fitness landscape is shown in Fig.\ 3C, in which successive population bifurcations lead to multiple speciation events in which the number of distinct population modes repeatedly doubles at an accelerating rate. In this sense, while Fig.\ 3A shows directional selection, and Fig.\ 3B shows disruptive selection, we can interpret Fig.\ 3C as a cascade of disruptive speciation events. We note that a wrapped domain is not necessary to produce oscillation; a population can oscillate back and forth along a one-dimensional landscape of finite domain width as well (Fig.\ S1), as long as the population variance remains bounded.


When the fitness landscape spans multiple trait dimensions, our model allows for the emergence of oscillatory evolution in as few as two trait dimensions. In the simple scenario of a Gaussian population density evolving on an initially Gaussian resource landscape (Fig.\ 4), we observe several characteristic behaviors of our model: rapid initial gradient ascent at early times before any appreciable landscape modification occurs (Fig.\ 4A), gradual oscillations outward from the initially Gaussian landscape peak as new selection gradients are established (Fig.\ 4B), and complex long-run behavior as roughness in the landscape accumulates (Fig.\ 4C) including stochastic reversals between clockwise and counterclockwise motion. This oscillatory behavior is reminiscent of the self-excited motions of volatile drops on swellable sheets\cite{chakrabarti2020}. The drop causes the sheet to swell (slowly) creating a local hill, which leads to the drop moving away (quickly); the sheet then swells (slowly) at the new location of the drop, and shrinks elsewhere, causing the drop to return (fast), leading to a slow-fast oscillatory motion, similar to a population moving on a regenerating landscape. Likewise, one can consider the analogy of a walker on a Penrose staircase (Fig.\ 4D), where the population is continuously climbing its local selection gradient but nonetheless moves perpetually in wide loops as the landscape under it is reshaped by the very population that traverses it. 


The complexity of solutions to Eqs.\ (1-2) increases with the size and dimensionality of the occupied landscape.   To develop some intuition for this, we use a low-dimensional projection of the dynamics in terms of the moments of the environment and the population, which we take to be approximately quadratic and Gaussian, respectively (Fig.\ 5A). At leading order (see End Matter and SI), we find that the population mean $\hat x$ and covariance matrix $\Sigma$ evolve according to the equations
\begin{align}
	\dot{\hat x} \approx\Sigma\beta,\quad
	\dot\Sigma\approx\Sigma H\Sigma^\mathsf{T},
\end{align}
where $\beta$ and $H$ denote the local selection gradient vector and landscape Hessian matrix, respectively, at state $\hat x$. The first equation serves as an analog to Fisherian selection on multidimensional landscapes \cite{good2013fluctuations}, while the second expresses how the spread of population density adapts to local landscape curvature. 
We also derive evolution equations for the moments of the local landscape (see End Matter): the local fitness $\hat u$, local selection gradient $\beta$, and local curvature $H$, which jointly follow
\begin{align}
\delta_t\hat u\approx\beta^\mathsf{T}\Sigma\beta,\quad
\delta_t\beta\approx H\Sigma\beta,\quad
\delta_tH\approx \varepsilon\hat u\Sigma^{-1}\hat n,
\end{align}
where the scalar $\hat n=\smash{\left|2\pi\Sigma\right|^{-\tfrac12}}$ is the maximal population density. Here we define the linear differential operator $\delta_t=\tfrac{d}{dt}+\varepsilon(\hat n-\phi)$, and the parenthetical term in the operator represents the reshaping of the landscape and therefore scales with $\varepsilon$; highly concentrated populations ($\hat n>\phi$) deplete their local landscape faster than it can regenerate. The right-hand sides of the first two equations in (5) are due to motion of the population on the landscape and persist even  when $\varepsilon=0$, wherein the operator $\delta_t$ is equivalent to $\tfrac{d}{dt}$.
Lastly, we find that the average fitness satisfies (see End Matter)
\begin{align}
	\phi=\hat u+\tfrac12\textup{tr}H\Sigma.
\end{align}
The trace of $H\Sigma$ indicates the degree to which the local landscape acts as a fitness peak (negative trace, $\hat u>\phi$) or valley (positive trace, $\hat u<\phi$). This system (4-5)  quantifies the relationship between a population's position and variance in trait space with its local fitness, local selection gradient, and local landscape curvature,  extending the Lande equation to account for non-static population variance ($\dot\Sigma\neq0$) and a non-static landscape ($\varepsilon\neq0$). In particular, we see that there are two effects of $\varepsilon$: mediating the reshaping of the landscape curvature $H$, and leading to an advection via the differential operator $\delta_t$ that governs the rate of landscape evolution as a whole.

In Fig.\ 5A, we show how the local fitness, its gradient, and the landscape curvature couple to the dynamics of the population, and in Fig.\ 6 (see End Matter), we show the form of the resulting dynamical equations. In Fig.\ 5B, we compare the results of solving the reduced order model (4-5) with simulations of the full model (1-2) for $\varepsilon \approx 0.03$; we see that the reduced order model is able to capture the essence of the full model, with high coefficients of determination ($R^2>0.90$) after elastic registration for each of the five moment dynamics (see SI) \cite{srivastava2011}.

\begin{figure}[b!]
	\centering
    \vspace*{-.1in}
	\includegraphics[width=\linewidth]{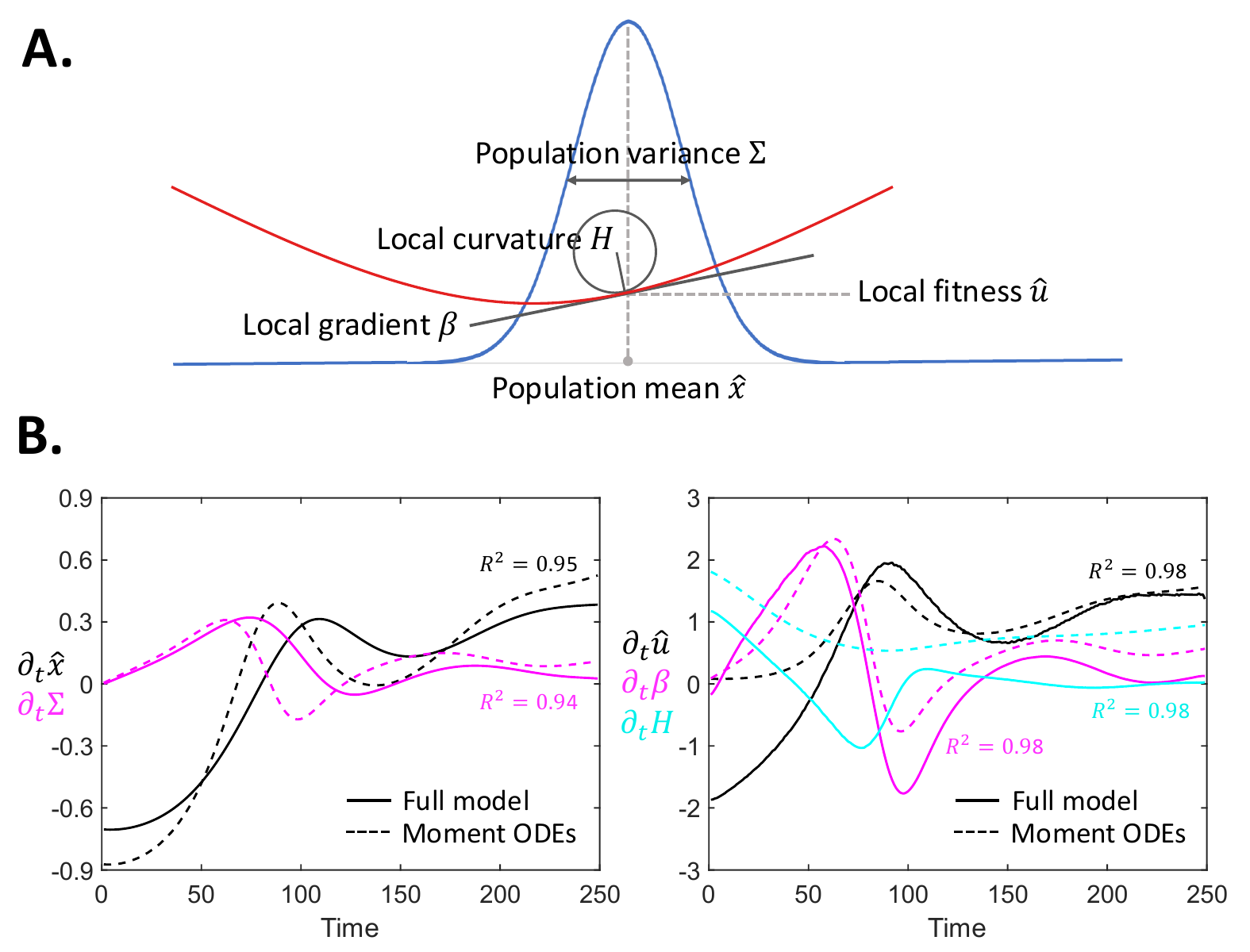}
    
    \vspace*{-.08in}
	\caption{\textbf{Local moment dynamics evolve as a system for an adapting population and its local landscape.} (A) Depicted here in a one-dimensional slice of trait space, a Gaussian population density profile with mean trait vector $\hat x$ and covariance matrix $\Sigma$ evolves on a paraboloidal landscape defined at $\hat x$ by its height $\hat u$ (local fitness), slope $\beta$ (selection gradient), and curvature $H$ (Hessian matrix). (B) The moments of solutions to the full model Eqs.\ (1-2) (solid) show high $R^2$ goodness of fit to numerical solutions of the approximate system of moment dynamics Eqs.\ (4-6) (dotted) after registration, with $\varepsilon=0.03$, initialized as the profiles from (A).}
\end{figure}

Several model assumptions could be relaxed. Possible extensions include multiple competing and co-evolving populations occupying disjoint or overlapping regions of trait space (see SI), environments with deterministic or stochastic temporal fluctuations that produce dynamic “seascapes” even without local populations \citep{Seascape1, Seascape2}, and diffusion in trait space driven by mutation or resource spreading \cite{Petak2025}. We also treat landscape sensitivity $\varepsilon$ as constant, though allowing dependence on time $t$ or trait $\mathbf{x}$ could reveal time-varying or anisotropic effects distinct from models where the fitness mapping $u$ is itself explicitly time- or trait-dependent \citep{mcgill2007evolutionary,metz1995adaptive}.

As our dynamics operate in a general trait space encompassing either discrete or continuous traits, spatial position for sessile organisms could be appended as additional trait dimensions. A spatial extension to our model might involve distance-decaying competition kernels or replacing global fitness with local averages (see SI), potentially yielding spatial niche coexistence, though we note also that diversification can arise even without niches \cite{Mahadevan2023}. Beyond populations reshaping their own landscape, exogenous perturbations could be applied to guide populations toward target traits \cite{krishnan2026}. 

\noindent \textbf{Acknowledgments}.  This work was supported by the Schmidt Science Fellowship (to A.H.) and NSF BioMatter Grant DMR-1922321, the Simons Foundation and the Henri Seydoux Fund (to L.M.). 
 
	\renewcommand{\bibname}{References}
	\bibliography{references}

%
%

\vspace*{.46in}

\subsection{END MATTER}

In the main text, in equations (1-2), substituting $\varepsilon=0$ yields the classical replicator equation, whereby the average fitness rate of change $\dot\phi=\sigma^2=\langle n,(u-\phi)^2\rangle$ represents the variance in fitness in the population (and is always non-negative); this is the Lande equation. However, if $\varepsilon\neq0$, then $u$ varies with time, and 
\begin{align}
\begin{split}
	\dot\phi&=\int_X \dot nu\, \textup d\mathbf x+\int_X n\dot u\, \textup d\mathbf x \\
	&=\int_X \dot n(u-\phi)\, \textup d\mathbf x
	+\int_X (n-\phi)\dot u\,\, \textup d\mathbf x \\
	&=\int_X n(u-\phi)^2 \textup d\mathbf x
	-\varepsilon\int_X (n-\phi)^2u\, \textup d\mathbf x \\
	&=\sigma^2-\varepsilon\omega^2,
\end{split}
\end{align}
where $\omega^2=\langle(n-\phi)^2,u\rangle$ represents the variance in occupation among resources on the landscape. 
Hence on a dynamic fitness landscape, the average fitness $\phi$ will decrease when and only when $\varepsilon>(\sigma/\omega)^2$.



To derive the rate of evolution $R$ in terms of our sole parameter $\varepsilon$, we define the non-dimensional time $T=Rt$ and consider the scaled version of (1-2) by writing $N=\psi n$ and $U=\psi'u$ where $\psi,\psi'$ are scaling factors. Taking $\psi'=\psi^{-1}$ to ensure $\langle N,U\rangle=\phi$, a rescaled version of (1-2) can be written as 
\begin{align}
	\frac{dN}{dT}&=\frac{\psi}{R} N(U-\phi_N), \\
	\frac{dU}{dT}&=-\frac{\varepsilon}{\psi R} U(N-\phi_U),
\end{align}
where $\phi_N=\phi/\psi$ and $\phi_U=\phi/\psi'$. In non-dimensional coordinates, setting the leading coefficients of each equation to unity yields $R \sim \psi$ and $\varepsilon \sim\psi R = R^2$, so that $R \sim \varepsilon^{1/2}$, the geometric mean of the leading coefficients in the original, dimensional system.


At each point in time, the state of the population in trait space can be summarized by its first and second moments: its mean position as a column vector $\hat x=\langle n,x\rangle$ and its covariance matrix $\Sigma=\langle n,(x-\hat x)(x-\hat x)^\mathsf{T}\rangle$. Likewise, the properties of the fitness landscape acting on this population can be summarized by its moments as calculated locally at the position $\hat x$ of the population: the local fitness $\hat u=u(x=\hat x)$, the local selection gradient $\beta=\nabla u(x=\hat x)$, and the local curvature as given by the Hessian matrix $H=\nabla^\mathsf{T}\nabla u(x=\hat x)$. 

To understand how these five key summary metrics interact in the context of our model, we consider the special case of a normally-distributed population evolving on a parabolic landscape:
\begin{align}
	n&\approx\hat n\exp\{-\tfrac12(x-\hat x)^\mathsf{T}\Sigma^{-1}(x-\hat x)\},\\
	u&\approx\hat u+\beta^\mathsf{T}(x-\hat x)+\tfrac12(x-\hat x)^\mathsf{T}H(x-\hat x),
\end{align}
where $\hat n=\smash{\left|(2\pi)^k\Sigma\right|^{-\tfrac12}}$ is the maximal population density in a $k$-dimensional trait space. The inner product of this Gaussian and paraboloid pair give the average fitness
\begin{align*}
	\phi=\langle n,u\rangle=\hat u+\tfrac12\textup{tr}H\Sigma.
\end{align*}
The trace of $H\Sigma$ indicates the degree to which the local landscape acts like a fitness peak (negative trace) or a fitness valley (positive trace); as expected, on peaks $\hat u\geq\phi$ while in valleys $\hat u\leq\phi$. At every instant in time, we squelch deviations from normality/parabolicity, but we allow all five summary metrics to evolve freely.

To determine the dynamics of this projection of our model dynamics into a five-dimensional state space, using the operator $\partial_t$ to denote differentiation in time, we find
\begin{align}
	\partial_t\hat x=\langle \dot n,x\rangle=\langle n(u-\phi),x- \hat x \rangle = \Sigma\beta,
\end{align}
where we have used $u-\phi \approx \beta^T (x-\hat x)$ and the definition of the covariance.  In accordance with the Lande equation, this can be interpreted as existing variation $\Sigma$ being acted upon by a local selection gradient $\beta$ to produce a directional rate of evolutionary change $\partial_t\hat x$ as the result of natural selection. Similarly, we can describe the evolution of the population variance by starting with $\Sigma=\int (x-\hat x) (x-\hat x)^T dx$, differentiating both sides, and inserting the quadratic expression for $u$ to arrive at 
\begin{align}
	\partial_t\Sigma=\Sigma H\Sigma^T,
\end{align}
where the local Hessian matrix $H$ describes the landscape curvature. 
If $H$ is positive definite, e.g.\ near a landscape valley, the population will expand outward to escape, while if $H$ is negative definite, e.g.\ near a landscape peak, the population will contract and localize about that peak.

In the case of a static landscape ($\varepsilon=0$), the local fitness $\hat u$ and selection gradient $\beta$ change only due to motion across the fitness landscape at the directional rate $\partial_t\hat x$, and we find
\begin{align}
	\partial_t\hat u&=\nabla u(\hat x)^T \partial_t \hat x =\beta^T\partial_t\hat x=\beta^T\Sigma\beta, \\
	\partial_t\beta&=\partial_t (\nabla u(\hat x(t))=[\nabla^T\nabla u(\hat x)] \partial_t \hat x=H\Sigma\beta,
\end{align}
where we have used (12) to derive both equations. The closure assumes that the population stays approximately Gaussian, meaning skewness and kurtosis are negligible. This holds when selection gradients change slowly across trait space, and the population variance is small enough that the Taylor expansion of $u$ holds locally, or more technically, when $\varepsilon \ll \Sigma^{-1/2}$. Breakdown occurs when branching happens or multiple modes form (as seen in Fig. 4C).

In this scenario, the preceding system (12-15) may be explicitly solved in terms of the fundamental matrix $\Psi=[I-H\Sigma_0 t]^{-1}$ and yields:
\begin{gather}
	\Sigma=\Psi\Sigma_0,\,
	\beta=\Psi\beta_0,\,
	\hat x=\hat x_0+\Psi\Sigma_0\beta_0 t,\\
	\hat u=\hat u_0-\tfrac12\Psi\beta_0^\mathsf{T}(I+\Psi^{-1})\Sigma_0\Psi\beta_0t,
\end{gather}
where the $\cdot_0$ subscript denotes initial conditions.
If and only if $H$ is negative definite, these moments converge to steady-state values in the long run: as $t\to\infty$, we obtain
\begin{gather}
	\Sigma\to0,\,
	\beta\to0,\\
	\hat x\to \hat x_0-H^{-1}\beta_0,\\
	\hat u\to\hat u_0-\tfrac12\beta_0^\mathsf{T}H^{-1}\beta_0,
\end{gather}
corresponding to a population with no variance, with optimal traits, and with maximal fitness. However, we note that the moment approximation is most effective over short time scales; over longer time scales, deviations from these predicted long-run behaviors accumulate over time.

Finally, we may ask how the curvature of the landscape evolves. To determine the dynamics of $H$, we take spatial second derivatives of the equation $\dot{u}=-\varepsilon u(n-\phi)$ at the population center $\hat{x}$. Under the Gaussian–parabolic closure, $n(\hat{x})=\hat n$, $\nabla n(\hat{x})=0$, and $\nabla\nabla^T n(\hat{x})=-\hat{n}\Sigma^{-1}$, while $u(\hat{x})=\hat{u}$, $\nabla u(\hat x)=\beta$, and $\nabla\nabla^T u(\hat{x})=H$. Using the product rule,
\begin{align}
\nabla\nabla^T(un) &= u\,\nabla\nabla^T n + 2\nabla u\nabla n+n\,\nabla\nabla^T u  \\
&= \hat{u}(-\hat{n}\Sigma^{-1}) + \hat{n}H \\
&= \hat{n}(H-\hat{u}\Sigma^{-1})
\end{align} at $x=\hat x$. From this, we obtain the evolution equation
\begin{equation}
\dot H=\nabla\nabla^T\dot{u} = -\varepsilon(\hat{n}-\phi)H + \varepsilon\hat{n}\hat{u}\Sigma^{-1}.
\end{equation}

\begin{figure}[!t]
	\centering
	\includegraphics[width=\linewidth]{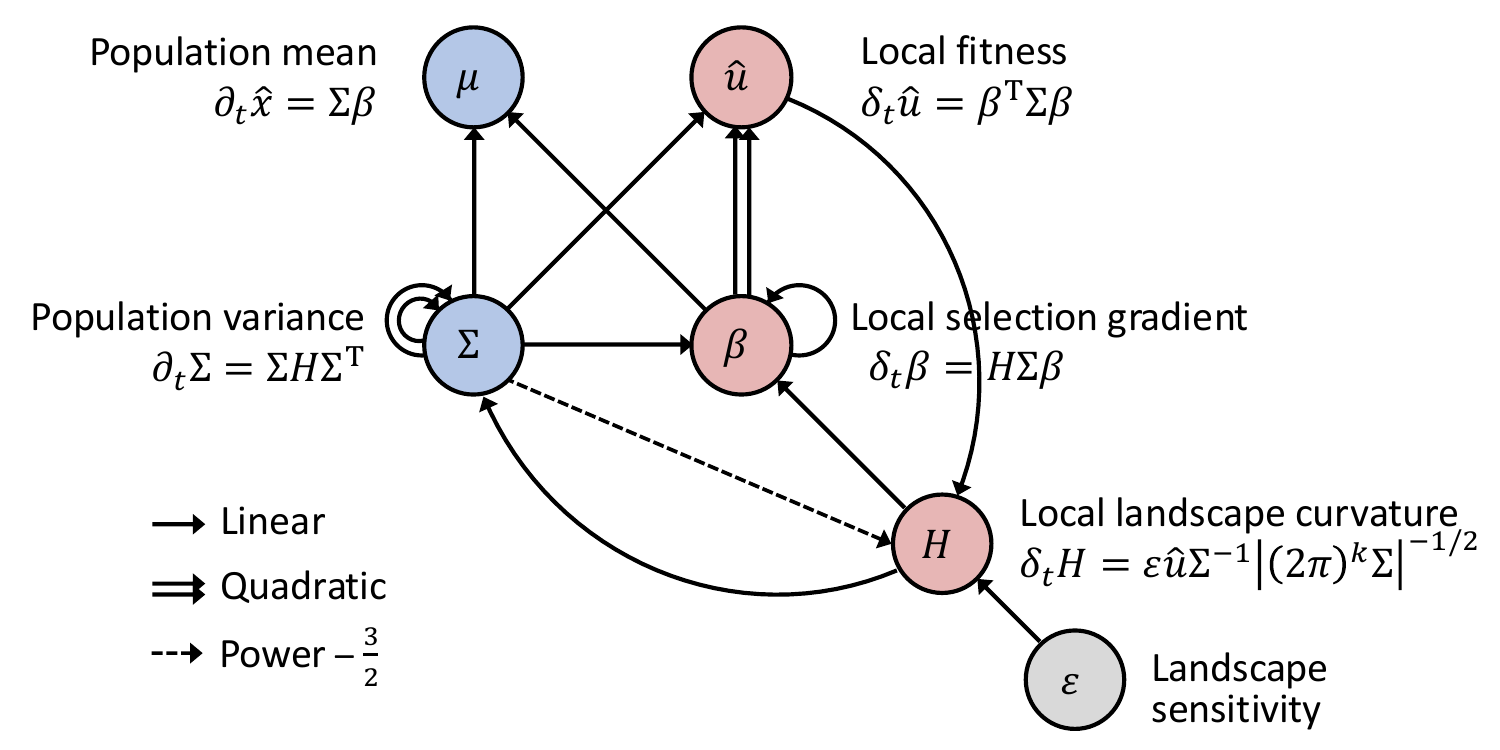}
	\caption{\textbf{Interaction network for local moment dynamics.} The dynamical equations (Eqs.\ 4-6) for the locally defined moments of the population (blue) and landscape (red) interact in the network shown here. Single, double, or dashed arrows respectively depict linear, quadratic, or power $-3/2$ dependence in the dynamics.}
\end{figure}

Because the curvature is evaluated at a moving point $\hat{x}(t)$, we introduce the advective derivative
$\delta_t := \frac{d}{dt} + \varepsilon(\hat{n}-\phi)$,
so that $\delta_t H = \nabla\nabla^T\dot{u}|_{x=\hat{x}}$. Substituting and simplifying yields
\begin{equation}
{\delta_t H = \varepsilon\hat{n}\hat{u}\Sigma^{-1},} \label{deltaH}
\end{equation}
and analogously, equations (14-15) become $\delta_t\hat u=\beta^T\Sigma\beta$ and $\delta_t\beta=H\Sigma\beta$.
 This represents local resource depletion that serves to increase $H$ over time. This local depletion effect scales with the landscape sensitivity $\varepsilon$, the amount of local resources that can be consumed $\hat u$, and local population concentration $\hat n$, as well as inversely with the population variance $\Sigma$ because a population with low variance will exert a more localized depletion effect that yields a lower radius of curvature and hence a greater curvature $H$. We note that these latter two influences (the height $\hat n$ and inverse variance $\Sigma^{-1}$ of the normally distributed population $n$) are negatively related because the total population density is normalized to remain constant over time.

 Taken in aggregate, this coupled system of moment equations can be visualized as an network in which the moments influence each other's dynamics (Fig.\ 6). This system accurately describes our model's dynamics over short time scales but gradually diverges from those dynamics over longer time horizons.
 
 We now turn to a precise analysis of this divergent long-term behavior, facilitated by introducing the population precision matrix $\Omega=\Sigma^{-1}$. Differentiating $\Omega$ in time gives the evolution equation
\begin{equation}
\dot\Omega=-\Sigma^{-1}\dot\Sigma\Sigma^{-1}=-\Sigma^{-1}(\Sigma H\Sigma)\Sigma^{-1}=-H.
\end{equation}
Applying the operator $\delta_t$ to both sides of this relation and substituting equation \ref{deltaH}, we obtain the damped matrix harmonic oscillator
\begin{align}
\ddot\Omega+\varepsilon(\hat n-\phi)\dot\Omega&=-\varepsilon\hat n\hat u\Omega.
\end{align}
This linearity here in $\Omega$ guarantees that chaotic long-term dynamics cannot be explained merely by these second-order moment approximations. Moreover, because $\varepsilon\hat n\hat u\Omega$ is positive definite, the approximating system will achieve $\Omega=0$ in finite time, meaning that the population covariance $\Sigma$ diverges.

Next, we differentiate the vector $v=\Omega\beta$ in time:
\begin{align}
\dot v&=\dot\Omega\beta+\Omega\dot\beta=-H\beta+\Omega H\Sigma\beta-\varepsilon(\hat n-\phi)\Omega\beta.
\end{align}
Because $\Omega=\Sigma^{-1}$, the first two terms precisely cancel, leaving only $\dot v=-\varepsilon(\hat n-\phi)v$; that is, $\delta_t v=0$. From this we conclude that $v$ never changes direction; we can write $v=\mu v_0$, where $\mu=\exp(-\int(\hat n-\phi)\,\d dt)$ is a non-negative, time-varying scalar and $v_0$ is a time-invariant direction in landscape space. Since $\beta=\Sigma v$, this implies that the population center moves with velocity
\begin{align}
\dot x=\Sigma\beta=\Sigma^2 v=\Sigma^2 \mu v_0.
\end{align}
Projecting this velocity onto the fixed direction $v_0$ gives $v_0^T\dot x=\mu\|\Sigma v_0\|^2\geq0$. From the non-negativity of $\mu$ and $\|\Sigma v_0\|^2$, we conclude that the population moves monotonically into the half-space defined by the vector $v_0$ at every point in time $t$. Therefore, the oscillations that we observe in our model---whereby the population trajectory can cycle around a point within the landscape---rely on aspects of our model that go beyond the reduced second-order moment dynamics derived here. 

\end{document}